\documentclass[seceq]{ptptex}

\usepackage{graphicx}
\usepackage{url}
\title{%        %You can use \\ for explicit line-break.
Impact of the Great East Japan Earthquake on Hotel Industry in Pacific
Tohoku Prefectures
}

\subtitle{From spatio-temporal dependence of hotel
availability}    %Use this 
\author{%       %Use \scshape for the family name.
Aki-Hiro \textsc{Sato}%
}

\inst{%     %Affiliation, neglected when [addenda] or [errata].
Department of Applied Mathematics and Physics, Kyoto University \\
Yoshida Honcho, Sakyo-ku, Kyoto 606-8501 Japan
}

\abst{%         %This abstract is neglected when [addenda] or [errata].
This paper investigates the impact of the Great Japan Earthquake
(and subsequent tsunami turmoil) on socio-economic activities 
by using data on
hotel opportunities collected from an electronic hotel booking
service. A method to estimate both primary and secondary regional effects
of a natural disaster on human behavior is proposed. It is confirmed that
temporal variation in the regional share of available hotels 
before and after a natural disaster may be an indicator to measure the
socio-economic impact at each district.
}

\begin{document}

\maketitle

\section{Introduction}
Since people in the world are also products of nature, the physical 
effects of natural environment on our society are remarkable. 
Specifically, natural disasters often affect our societies
significantly. Therefore, we need to understand the subsequent impact of
natural disasters on human behavior from both economical and social
perspectives.

The first Great East Japan Earthquake hit at 14:46 on 11 March 2011 in
Japanese local time (05:46 in UTC). Within 20 minutes after shaking, 
huge tsunamis had devastated cities along Japan's northeastern
coastline. In fact, 
destruction was largely physical, but social infrastructures have been
damaged. Currently, it is considerably significant for us to
understand its subsequent impact on our socio-economic activities.

According to Sigma 2/2011 from Swiss Re, significant insurance losses
are expected from the 11 March Tohoku Earthquake (and subsequent
tsunami), which resulted in the death and disappearance of more than
23,000 people~\cite{Sigma-2-2011}. Economic activities before and after the
Great East Japan disaster completely changed due to the physical
destruction observed in many industrial sectors. The fiscal cost of this
disaster to Japan has been conservatively estimated at \$200
billion. Actually, the short-term impact on Japanese growth is likely to
be negative and potentially quite large. Nevertheless, reconstruction
efforts are likely to get underway, which will provide a substantial boost
to growth by the end of this year.

Monitoring regional human behavior may provide decision-makers
with useful insights on the management of re-establishment after
disasters. However, in general, it is not so easy to collect
large-scale data on human activities before and after the natural disasters with
both a high level of detail and a large number of samples. In this case, it is
important to find an adequate proxy variable. 

Tourism demand is particularly sensitive to security and health
concerns. Estimation of the economic impact of changes
in the demand for tourism has typically been investigated with
several methods. Blake et al. analyzed the effects of crisis using a
computational general equilibrium model of the US and also examined
potential and actual policy responses to the crisis~\cite{Blake:03}.
Moreover, the problem of estimating demand from censored booking data
has been recognized for many years in the hotel industry. Patrick et
al. developed parametric regression models that consider not only the
demand distribution, but also the conditions under which the data were
collected~\cite{Patrick:02}. Sato investigated regional
patterns of Japanese travel behavior by using the EM algorithm for finite 
mixtures of Poisson distributions~\cite{Sato:11}.

Furthermore, understanding what motivations influence people's travel
habits and destination selections is crucial to predicting their future
travel patterns. A review of the literature on tourist motivation
reveals that a model that views motivations on the two dimensions of
``push'' and 
``pull'' factors has been generally accepted~\cite{Cha}. The idea behind
this two-dimensional approach is that people travel because they are pushed
by their own internal forces and pulled by the external forces of the
destination attributes~\cite{Dann:77}. The pull factors 
originate from the destination properties (supply). The push factors belong
to consumers (demand). More recently, Tkaczynski et al. applied
the stake-holder theory, a management theory proposed by
Freeman (1984)~\cite{Freeman}, to a destination in
tourism~\cite{Tkaczynski}. The existence of hotel accommodations implies that
pull factors are present in the district where they are located.

Therefore, we may assume that changes of demand and supply in
the hotel industry can reflect both the social and economic impact of natural
disasters on human behavior. In this article, the use of
data on hotel availability collected from an electronic hotel booking
service is suggested for this purpose. Hotel opportunities are
handled in real-time and the geographical coverage of hotel locations
seems high. Spatio-temporal dependence of hotel availability
can be analyzed from physical point of view.

According to a recent report by Japan Tourism Agency in the Ministry
of Land, Infrastructure, Transport and 
Tourism, the total number of accommodations, including hotels, Japanese
inns, and pensions all over Japan, is estimated as 53,468 during
2010~\cite{JTA}. Specifically, in the prefectures that have been damaged
by the earthquake and tsunami (Iwate, Miyagi, and 
Fukushima), it is estimated that there are 3,846 accommodation
facilities. Notably, 
hotels in this area are so densely located that we can use the data to
measure the socio-economic impact. Based upon this idea, the number of
available hotels before and after the Great East Japan Earthquake is
examined in this study. Hotel availability is assumed to be a proxy
variable of human mobility, so comparative analysis before and after the
disaster is conducted.

\section{Data and methods}
\label{sec:data}
We used data collected from a Japanese hotel booking site named 
Jalan via a Web API (Application Programing Interface)~\cite{Jalan}. The
Jalan is one of the most popular hotel reservation services in Japan.
The Web API is an interface code set that is designed to
simplify development of application programs.

In order to estimate both economic and social damages in three
Tohoku prefectures (Iwate, Miyagi, and Fukushima), we focus on the 
number of available hotels in each district before and after the Great
East Japan Earthquakes and Tsunami. We selected 21 districts in three
prefectures, as shown in Table \ref{tab:district} and two periods,
one before and one after the disaster.

%Fig. \ref{fig:graph} shows the daily numbers of available hotels in each
%district before and after the disaster. In these graphs we can see that
%the total number of available hotels decreased after the earthquake and
%that the numbers sharply decreased after the disaster. 

The data on hotels in this area cover about 31\% of the potential
hotels. Therefore, we have to estimate the uncensored states from these censored
booking data. If we assume that the hotels in the data are sampled from
uncensored data in a homogeneous way, then the relative frequency of the
available accommodations from censored data can approximate 
the true value, computed from uncensored data.

In order to conduct a quantitative study, let $x_i(t,s) \quad (i=1,\ldots,K;
t=1,\ldots, T)$ be the number of available hotels in district $i$ at day
$t$ in period $s$. Then the relative frequency at district $i$ can be
calculated as 
\begin{equation}
p_i(s) = \frac{\sum_{t=1}^T x_i(t,s)}{\sum_{i=1}^K \sum_{t=1}^T x_i(t,s)}.
\end{equation}
Let us consider a ratio of the relative frequencies after and before a
specific event
\begin{equation}
q_i(a;b) = p_i(a)/p_i(b),
\label{eq:ratio}
\end{equation}
where $p_i(a)$ and $p_i(b)$ represent the relative frequencies after and
before the event, respectively. Obviously, Eq. \ref{eq:ratio} can
be rewritten as
\begin{equation}
q_i(a;b) = \frac{n_i(a)}{n_i(b)}/\frac{N(a)}{N(b)},
\label{eq:change-of-ratio}
\end{equation}
by using $n_i(s)$ and $N(s)$, which are the number of available hotels
in district $i$ within the period $s$ and the total number at that
moment, respectively. Since
$N(a)/N(b)$ is independent of $i$, $q_i(a;b)$ should be proportional to
the ratio of the number of hotels after and before the event. 

\section{Analysis}
\label{sec:analysis}
The 3rd column in Table \ref{tab:district} shows $q_i(a;b)$, where the
term $b$ represents May 2010 (before the disaster), and the term $a$ May
2011 (after the disaster), respectively. Since the value of $q_i(a;b)$
is related to damages to hotels in district $i$, $q_i(a;b)<1$ implies that
available hotels decreased after the earthquake 
at $i$, compared to the ratio of the total number of hotels. Similarly
$q_i(a;b)>1$ means that they increased or maintained at $i$. 

From this it is estimated that Sanriku Kaigan, Oshu, Hiraizumi,
Ichinoseki, Sendai, Matsushima, Shiogama, Fukushima, Nihonmatsu, Soma,
Koriyama, Iwaki and Futaba were significantly damaged by the earthquake
and tsumani. We may assumed that the decrease of $q_i(a;b)$ at district
$i$ results from both a decrease in supply and an increase in
demand. The decrease in supply is caused by the physical destruction of
infrastructure in this case. The increase in demand comes from behavior
of individuals. The regional dependence of supply can be estimated from
the number of destroyed houses in each district. 

To do so, we calculated the numbers of
both completely destroyed and partially destroyed houses in each
district from the data downloaded from a Web page of National Research
Institute for Earth Science and Disaster Prevention~\cite{NDIS}. 
The numbers were calculated by summing the number of destroyed houses in
a town or city included in each district. Table \ref{tab:district} shows
the numbers of destroyed houses. From this histogram, it can be
seen that house damages were concentrated in the maritime area of these
prefectures.

We can confirm that the damages to houses were serious in
Sanrikukaigan, Sendai, Matsushima, Shiogama, Ishinomaki, Kesennuma,
Soma, Koriyama, Iwaki and Futaba. The highest number of
completely destroyed houses in Ishinomaki and Kesennuma,
with 33,661 homes destroyed. The second-highest is 21,789, in Sendai. 
The third-highest is 18,098, in Sanrikukaigan. The
highest number of partially destroyed houses is 37,522, in
Sendai. The second-highest is 17,614, in Iwaki and Futaba. 
The third-highest is 12,185, in Koriyama. 

From an official report by each Prefecture~\cite{Iwate,Miyagi,Fukushima}
the number of evacuees who have stayed at public refuge holes is counted
at each district. The 6th column in Table \ref{tab:district} shows the
number of evacuees as confirmed at 1st May 2011. The number of evacuees in each
district is proportional to the number of completely destroyed
houses. It is found that physical damages were not serious from
both the number of destructed houses and evacuees in Oushu, Hiraizumi,
Ichinoseki and Fukushima, Nihonmatsu. The degree of damages which is
measured by physical damages is consistent with a value of $q_i$
computed by the proposed method except these districts.

\begin{figure}[h]
\centering
\includegraphics[scale=0.8]{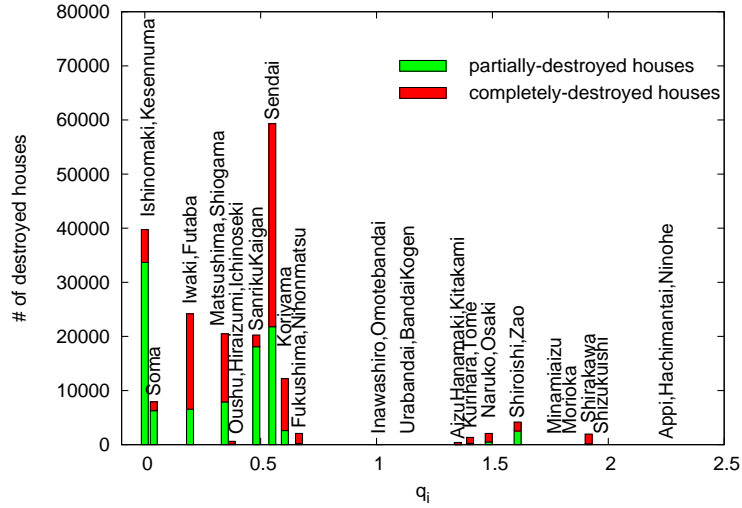}
\caption{Relationship between the ratio $q_i(a;b)$ before
and after the Great East Japan Earthquake, and the number of destroyed
 houses, as confirmed at the end of September 2011.}
\label{fig:corr}
\end{figure}

The Figure \ref{fig:corr} shows correlation among the number of
destroyed houses and $q_i$. In fact, where the ratio $q_i(a;b)$ is
greater than 1, the number of destroyed houses is not significant.
We confirmed that the ratio $q_i(a;b)$ may measure the degree of damage
to economic activities through the travel industry. However, it was not
confirmed that the significant physical damages to houses in Oushu,
Hiraizumi, Ichinoseki, Fukushima, and Nihonmatsu even have a ratio less
than 1. It may be thought that hotels in Oushu, Hiraizumi, and
Ichinoseki were used by works and evacuated victims of the disaster. The
decrease in available hotels in Fukushima and Nihonmatsu may be 
related to the accidents at the Fukushima Daiich nuclear power plant. 

The number of evacuated victims of the disaster in each
prefecture according to an official announcement by Japanese Cabinet
Office on 3 June 2011 is shown in Table \ref{tab:victims}. In the case
of the Fukushima prefecture, 17,874 evacuated people were in hotels at
that moment. Although we could not collect correct data on hotel
booking on these two districts, we may assume that hotels in these
districts were occupied by individuals (refugees, workers, volunteers
and civil groups). 

Consequently, we concluded that the physical damage 
to buildings constructions is estimated to be small if the ratio is
larger than 1. Moreover, it is found that regional distribution of
evacuated victims and workers may influence the ratio. This 
represents the secondary effects of natural disasters on human
behavior.

%\begin{figure}[h]
%\centering
%\includegraphics[scale=0.95]{../prog/hotelcountbefore.eps}(a) 
%\includegraphics[scale=0.95]{../prog/hotelcountafter.eps}(b)
%\caption{(Color Online) The numbers of available hotels during a period
% from 1st to 31st May 2010 (a) and 1st to 31st May 2011 (b).}
%\label{fig:graph}
%\end{figure}

%\begin{figure}[hbt]
%\centering
%\includegraphics[scale=0.95]{../prog/prob.eps}(a)
%\includegraphics[scale=0.95]{completely-destroyed.eps}(b)
%\caption{(a) The ratios of relative frequencies after the Great East Japan
% Earthquake (May 2010) to those before it (May 2011). (b) The numbers of
% both completely-destroyed and partially-destroyed houses
% confirmed at the end of August 2011.}
%\label{fig:prob}
%\end{figure}

\begin{table}[h]
\caption{The ratio of the number of available hotels during the period from
1st to 31st May 2011 to that during the period from 1st to 31st May 2010
(after and before the Great East Japan Earthquake), the number of
 both completely destroyed houses and partially destroyed houses, as
 confirmed at the end of September 2011 and the number of evacuees, as
 confirmed at 1st May 2011.}
\label{tab:district}
\centering
\begin{tabular}{lllllll}
\hline
prefecture & district & ratio & complete collapse & partial collapse &
 evacuees \\
\hline
Iwate & Shizukuishi & 1.970 & 0 & 0 & 372 \\
      & Morioka & 1.834 & 0 & 4 & 366 \\
      & Appi,Hachimantai,Ninohe & 2.250 & 3 & 0 & 0 \\
      & Hanamaki,Kitakami,Tohno & 1.350 & 27 & 364 & 853 \\
      & SanrikuKaigan & 0.481 & 18,098 & 2,166 & 12,896 \\
      & Oushu,Hiraizumi,Ichinoseki & 0.374 & 83 & 533 & 338 \\
\hline
Miyagi & Sendai & 0.550 & 21,789 & 37,522 & 3,608 \\
       & Matsushima,Shiogama & 0.345 & 7,895 & 12,581 & 5,115 \\
       & Ishinomaki,Kesennuma & 0.0 & 33,661 & 6,083 & 23,840 \\
       & Naruko,Osaki & 1.484 & 486 & 1,577 & 929 \\
       & Kurihara,Tome & 1.404 & 224 & 1,105 & 1,049 \\
       & Shiroishi,Zao & 1.608 & 2,522 & 1,644 & 1,612 \\
\hline
Fukushima & Fukushima,Nihonmatsu & 0.665 & 168 & 1,898 & 1,321 \\
          & Soma & 0.038 & 6,279 & 1,618 & 1,969 \\
          & Urabandai,BandaiKogen & 1.134 & 0 & 0 & 2 \\
          & Inawashiro,Omotebandai & 1.009 & 10 & 12 & 303 \\
          & Aizu & 1.352 & 4 & 27 & 266 \\
          & Minamiaizu & 1.768 & 0 & 0 & 14 \\
          & Koriyama & 0.604 & 2,596 & 12,185 & 2,489 \\
          & Shirakawa & 1.915 & 135 & 1,820 & 418 \\
          & Iwaki,Futaba & 0.195 & 6,550 & 17,614 & 2,115\\
\hline
\end{tabular}
\end{table}

\begin{table}[h]
\caption{The number of evacuees of the Great East Japan Earthquake at
 three prefectures (Iwate, Miyagi, and Fukushima). The data were officially
 announced by the Japanese Cabinet Office on 3rd June 2011.}
\label{tab:victims}
\centering
\begin{tabular}{lllll}
\hline
prefecture & A: public places & B: hotels & C: others & A+B+C \\
\hline
Aomori & 0 & 78 & 777 & 855 \\
Iwate & 9,039 & 2,007 & 14,701 & 25,747 \\
Miyagi & 23,454 & 2,035 & - & 25,489 \\
Akita & 128 & 619 & 909 & 1,656 \\
Yamagata & 305 & 779 & 2,366 & 3,450 \\
Fukushima & 6,105 & 17,874 & - & 23,979 \\
\hline
\end{tabular}
%\begin{tabular}{lll}
%\hline
%prefecture & district & the number of evacuated victims \\
%Iwate & Shizukuishi & 372 \\
%      & Morioka  & 366 \\
%      & Appi,Hachimantai,Ninohe & 0 \\
%      & Hanamaki,Kitakami,Tohno & 853 \\
%      & SanrikuKaigan & 35814 \\
%      & Oushu,Hiraizumi,Ichinoseki & 338 \\
%\hline
%Miyagi & Sendai & 3608 \\
%       & Matsushima,Shiogama & 5115 \\
%       & Ishinomaki,Kesennuma & 23840 \\
%       & Naruko,Osaki & 929 \\
%       & Kurihara,Tome & 1049 \\
%       & Shiroishi,Zao & 1612 \\
%\hline
%Fukushima & Fukushima,Nihonmatsu & 1321 \\
%          & Soma & 1969 \\
%          & Urabandai,BandaiKogen & 2 \\
%          & Inawashiro,Omotebandai & 303 \\
%          & Aizu & 266 \\
%          & Minamiaizu & 14 \\
%          & Koriyama & 2489 \\
%          & Shirakawa & 418 \\
%          & Iwaki,Futaba & 2115 \\
%\hline
%\end{tabular}
\end{table}

\section{Transient behavior}
\begin{figure}[h]
\centering
\includegraphics[scale=0.9]{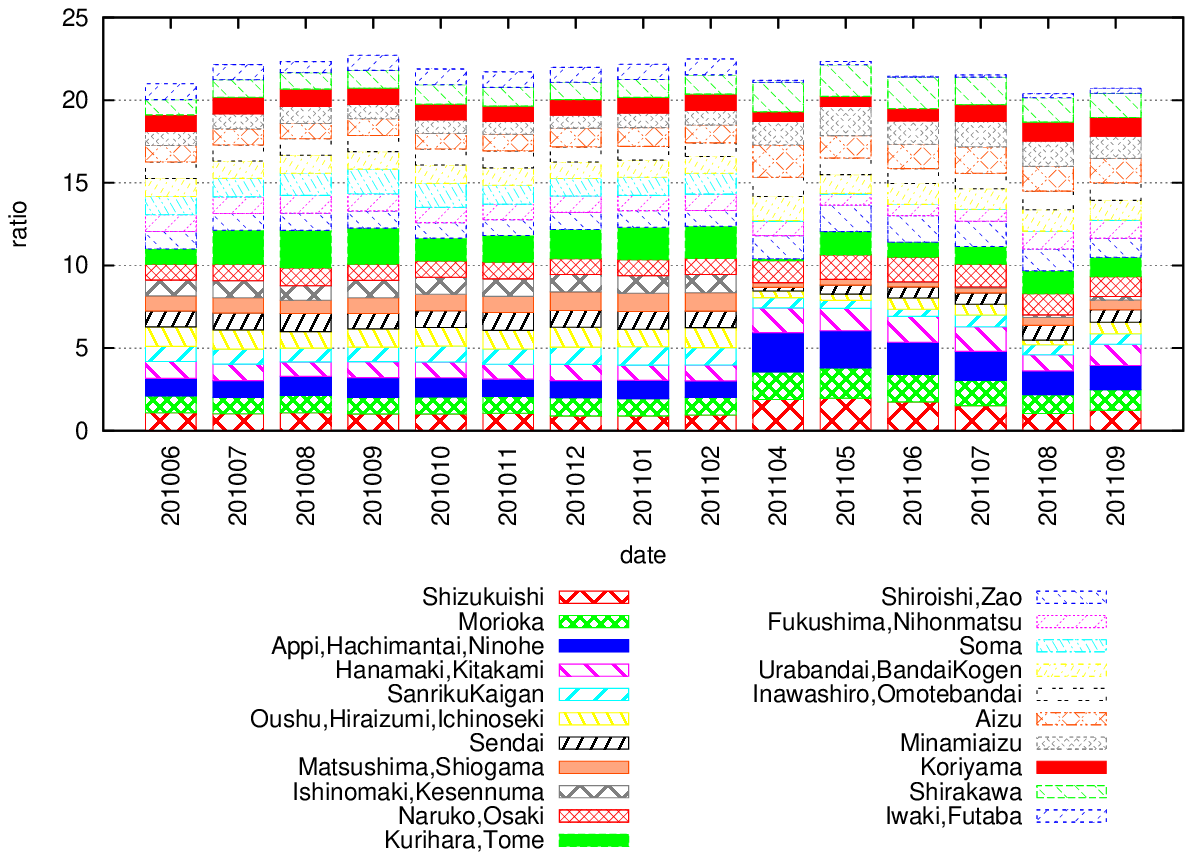}
\caption{(Color online) The ratios of the relative frequencies of
 available hotels from May 2010 to September 2011. The base period is
 fixed as May 2010.} 
\label{fig:ratios}
\includegraphics[scale=0.9]{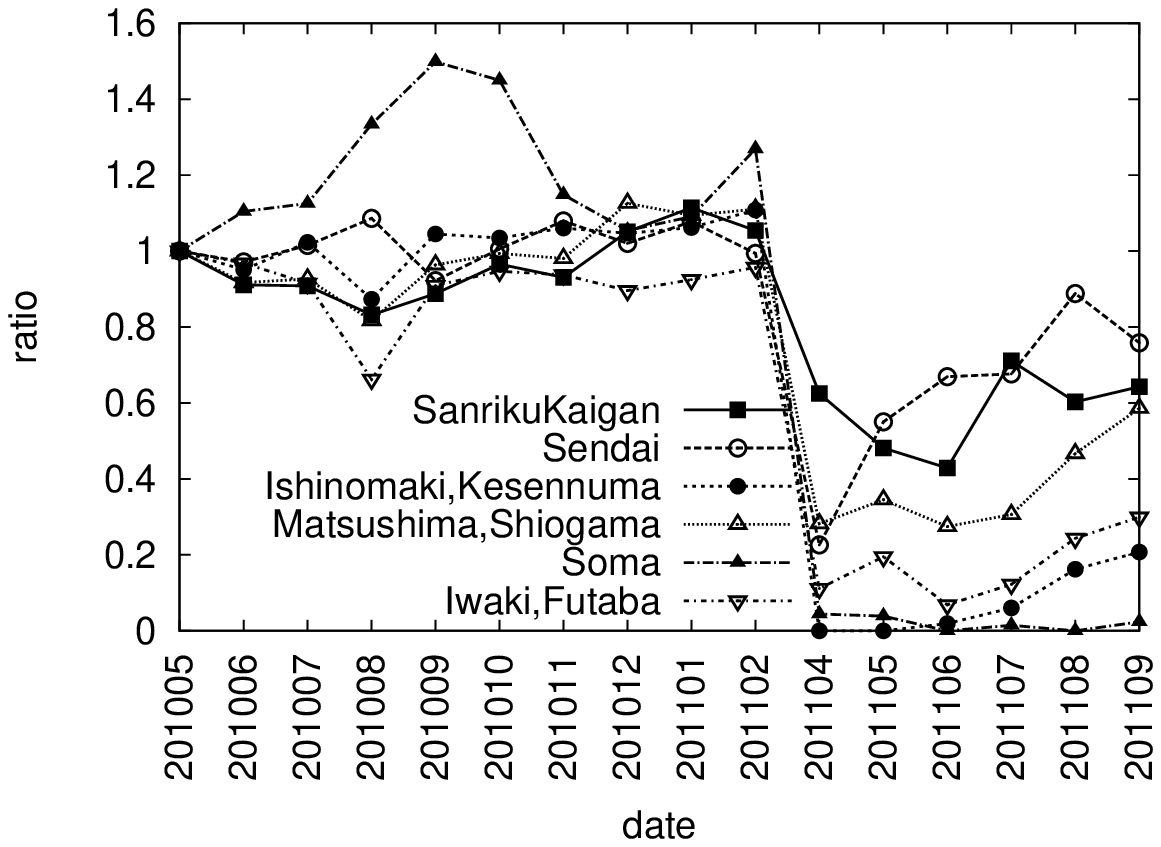}
\caption{The ratios of the relative frequencies of available hotels from
 May 2010 to September 2011 at several districts (SanrikuKaigan, Sendai,
 Ishinomaki, Kesennuma, Matsushima, Shiogama, Soma, Iwaki and
 Futaba). The base period is fixed as May 2010.}
\label{fig:ratios2}
\end{figure}

In this section, we investigate temporal development of hotel availability
before and after the Great East Japan Earthquake. By using
Eqs. (\ref{eq:ratio}) and (\ref{eq:change-of-ratio}), we compute
$q_i(a;b)$ on the basis of the regional share of available hotels in May
2010. Fig. \ref{fig:ratios} shows $q_i(a;b)$, where $b$
represents May 2010, and $a$ the period from May 2010 to September 2011. It
is found that the ratios are stable and take values around 1 at 
all the districts except Kurihara and Tome until February 2011.
 
However, after the earthquake (from April 2011) the ratios drastically
changed. Fig. \ref{fig:ratios2} shows the ratios of several
districts. The ratio in April 2011 reflects the degree of physical
damage just after the earthquake, as shown in Section
\ref{sec:analysis}. Therefore, the temporal variation can be assumed 
to show the degree of both physical and social damage. 

In Fig. \ref{fig:ratios2}, we see that 22.6\% of the hotels in Sendai 
had recovered by April 2011, 55\% by May 2011, 66.9\% by June 2011,
67.6\% by July, 88.8\% by August 2011 and 75.8\% by September 2011, in
comparison with May 2010. Hotels along the coastlines (Sanrikukaigan,
Matsushima, Shiogama, Ishinomaki, Kesennuma and Iwaki) are eventually
recovering, though more slowly than at Sendai. In the case of Ishinomaki
and Kesennuma, the hotels completely disappeared in April 2011, but 20.1\%
of the hotels had recovered by September 2011. In contract to these areas,
hotels at Soma had not recovered by the time this text was written, in
September 2011.

\section{Conclusion}
A method to measure changes in human activities before and after a natural
disaster from both social and economic perspectives with data
collected from an electronic booking site was proposed. Regional shares of
the number of available hotels were introduced and a ratio of the share
after the disaster to that before it was defined. Hotel
availability before and after the Great East Japan Earthquake and
Tsunami was analyzed through a comparison of the number of destroyed houses with
the ratio. As a result, we confirmed that the physical damage 
to buildings constructions is estimated to be small if the ratio is
larger than 1. Moreover, it is found that regional distribution of
evacuated victims and workers may influence the ratio. This 
represents the secondary effects of natural disasters on human
behavior.

\section*{Acknowledgements}
The author thanks the Yukawa Institute for Theoretical Physics at Kyoto
University. Discussions during the YITP workshop YITP-W-11-04 on 
``Econophysics 2011: The Hitchhiker's Guide to the Economy'' were useful
in completing this work. The author are thankful to Prof. Hideaki Aoyama
and Prof. Zdzis\l aw Burda for valuable suggestions. The author also
thanks Mr. Sigehiro Kato, Mr. Hiroshi Yoshimura, Mr. Kotaro Sasaki and
Ms. Yoko Miura of Recruit Cooperation for their stimulating discussions.
%\appendix
%\section{First Appendix} %Empty argument \section{} yields `Appendix'. 
%
%\section{Second Appendix}


\begin{thebibliography}{99}
%%%%%%%%%%%%%%%%%%%%%%%%%%%%%%%%%%%%%%%%%%%%%%%%%%%%%%%%%%%%%
% Some macros are available for the bibliography:
%  o for general use
%    \JL : general journals                 \andvol : Vol (Year) Page
%  o for individual journal 
%    \AJ   : Astrophys. J.           \NC         : Nuovo Cim.
%    \ANN  : Ann. of Phys.           \NPA, \NPB  : Nucl. Phys. [A,B]
%    \CMP  : Commun. Math. Phys.     \PLA, \PLB  : Phys. Lett. [A,B]
%    \IJMP : Int. J. Mod. Phys.      \PRA - \PRE : Phys. Rev. [A-E]     
%    \JHEP : J. High Energy Phys.    \PRL        : Phys. Rev. Lett.
%    \JMP  : J. Math. Phys.          \PRP        : Phys. Rep.
%    \JP   : J. of Phys.             \PTP        : Prog. Theor. Phys.     
%    \JPSJ : J. Phys. Soc. Jpn.      \PTPS       : Prog. Theor. Phys. Suppl.
% Usage:
%  \PRD{45,1990,345}          ==> Phys.~Rev.\ D \textbf{45} (1990), 345
%  \JL{Nature,418,2002,123}   ==> Nature \textbf{418} (2002), 123
%  \andvol{123,1995,1020}    ==> \textbf{123} (1995), 1020
%%%%%%%%%%%%%%%%%%%%%%%%%%%%%%%%%%%%%%%%%%%%%%%%%%%%%%%%%%%%%
\bibitem{Sigma-2-2011} Sigma 2/2011, Swiss
	Re. \url{http:www.swissre.com/publications/}.
\bibitem{Blake:03} %Adam Blake, M.Thea Sinclair, TOURISM CRISIS MANAGEMENT:
%	US Response to September 11, 
	\JL{Annals of Tourism Research, 30, 2003, 813}%--832.
\bibitem{Patrick:02} %Liu, Patrick H.;  Smith, Stuart;   Orkin, Eric B.;
	%Carey, George, "Estimating unconstrained hotel demand based on
	%censored booking data", 
	\JL{Journal of Revenue and Pricing Management,1, 2002, 1, 121}%--139.
\bibitem{Cha} %Sukbin Cha, Ken W. McCleary, and Muzaffer Uysal, ``Travel
	%Motivations of Japanese Overseas Travelers: A Factor-Cluster
	%Segmentation Approach'', 
	\JL{Journal of Travel Research, 34, 1995, 33}%-39. 
\bibitem{Sato:11} %Aki-Hiro Sato, ``Patterns of regional travel behavior: An
	%analysis of Japanese hotel reservation data'', 
	International Review of Financial Analysis (2011) in press 
\bibitem{Dann:77} %G. Dann, "Anomie, Ego-Enhancement and Tourism", 
	\JL{Annals of Tourism Research, 4, 1977, 184}%-194. 
\bibitem{Tkaczynski} %Aaron Tkaczynski, Sharyn Rundle-Thiele, and Narelle
%       Beaumont, "Destination Segmentation: A Recommended Two-Step
%	Approach", 
	\JL{Journal of Travel Research, 49, 2010, 139}%-152.
\bibitem{Freeman} R.E. Freeman, \it{Strategic Management: A Stakeholder
	Approach} (Boston, Pitman, 1984)
\bibitem{JTA} The data is downloaded from a Web page of Japan Tourism
	Agency in Ministry of Land, Infrastructure, Transport and
	Tourism: \url{http://www.mlit.go.jp/kankocho/siryou/toukei/index.html}
	(31 Aug 2011).
\bibitem{Iwate} The data is downloaded from a Web page of Iwate
	Prefecture: \url{http://www.pref.iwate.jp/~bousai/taioujoukyou/201105011700hinanbasyo.pdf}
\bibitem{Miyagi} The data is downloaded from a Web page of Miyagi
	Prefecture: \url{http://www.pref.miyagi.jp/kikitaisaku/higasinihondaisinsai/pdf/5011900.pdf}
\bibitem{Fukushima} The data is downloaded from a Web page of Fukushima
	Prefecture: \url{http://www.pref.fukushima.jp/j/hinanjolist0501.pdf}
\bibitem{Jalan} The data is collected from Jalan Web Service:
	\url{http://www.jalan.net/}.
\bibitem{NDIS} The data is downloaded from a Web page of National Research
	Institute for Earth Science and Disaster Prevention:
	\url{http://www.j-risq.bosai.go.jp/ndis/} (31 Aug 2011).
\end{thebibliography}
\end{document}